          [2001/04/25 1.1 (PWD)]
\documentclass[a4paper,twocolumn]{esapub} 
\usepackage{natbib}
\usepackage{graphicx}

\long\def\unmarkedfootnote#1{{\long\def\@makefntext##1{##1}\footnotetext{#1}}}

\title{ Hypernovae as possible sources of Galactic positrons}
\author{St\'ephane Schanne$^{(1,0)}$}
\author{Michel Cass\'e$^{(1,2)}$}
\author{Bertrand Cordier$^{(1)}$}
\author{Jacques Paul$^{(1,3)}$}
\affil{$^{1}$ CEA-Saclay, DSM/DAPNIA/Service d'Astrophysique, 91191 Gif-sur-Yvette, France }

\begin{document}


\maketitle

\begin{abstract}

	INTEGRAL/SPI\unmarkedfootnote{
corresponding author: schanne@hep.saclay.cea.fr \\
$^{2}$ Institut d'Astrophysique de Paris, 98 bis Boulevard Arago, 75014 Paris, France \\
$^{3}$ F\'ed\'eration de Recherche Astroparticule et Cosmologie, 
Coll\`ege de France, 11 Place Marcellin Berthelot, 75231 Paris, France
}
 has recently observed a strong and extended emission resulting from electron-positron annihilation located in the Galactic center region, consistent with the Galactic bulge geometry, without a high energy gamma-ray counterpart, nor in the 1809 keV $^{26}$Al decay line.
In order to explain the rate of positron injection in the Galactic bulge, estimated to more than 10$^{43}$ s$^{-1}$, the most commonly considered positron injection sources are type Ia supernovae. However, SN~Ia rate estimations show that those sources fall short of explaining the observed positron production rate, raising a challenging question about the nature of the Galactic positron source. In this context, a possible source of Galactic positrons could be supernova events of a new type, as the recently observed SN2003dh/GRB030329, an exploding Wolf-Rayet star (type Ic supernova) associated with a hypernova/gamma-ray burst; the question about the rate of this kind of events remains open, but could be problematically low.
	In this paper, we explore the possibility of positron production and escape by such an event in the framework of an asymmetric model, in which a huge amount of $^{56}$Ni is ejected in a cone with a very high velocity; 
the ejected material becomes quickly transparent to positrons, which spread out in the interstellar medium.

\end{abstract}

\section{511 keV gamma-ray line emission from the Galactic center region}

	The spectrometer SPI \citep{schanne2002,attie2003,roques2003,vedrenne2003} on ESA's gamma-ray satellite INTEGRAL, in orbit since October 2002, has recently reported its first results on the observation towards the Galactic center region of the 511 keV gamma-ray line emission resulting from e$^{+}$ e$^{-}$ annihilation \citep{jean2003}.
	The flux of annihilation detected by SPI is $\Phi_{511}=$(0.99$^{+0.47}_{-0.21}$) $\times$ 10$^{-3}$ ph cm$^{-2}$ s$^{-1}$, is concentrated in a narrow gamma-ray line at an energy of 511.06$^{+0.17}_{-0.10}$ keV and whose intrinsic line width is evaluated to 2.95$^{+0.45}_{-0.51}$ keV (FWHM), while the instrumental resolution at this energy is 2.16 keV. 
	Furthermore, indications on the spatial shape of the 511 keV emission region have also been published by \cite{jean2003} and \cite{knodl2003}.
	A single point source located in the Galactic center region is excluded. 
	When fitting a spherically symmetric Gaussian distribution to the emission region, the best fit is obtained (with a significance level of 12 $\sigma$) for a Gaussian centered on the Galactic center, with an width of 10$^\circ$ (FWHM).
	With 95\% C.L. the emission region has an extension in the range between 6$^\circ$ and 18$^\circ$ (FWHM).
	A Richardson-Lucy deconvolution confirms this result.
	In addition, no significant emission from the Galactic disk has been yet detected by SPI.

	We know the Galactic center to be located at a distance of $R_{o}=$8.0$\pm$0.4 kpc with good accuracy, thanks to the precision obtained with a new geometric determination, a Kepler orbit fit to the star S2 orbiting the central black hole, by \cite{eisen2003}.
	If we assume that the 511 keV emission takes place near the Galactic center, we can conclude that the spatial size of the emission region is $D=$1.4 kpc in diameter, which coincides roughly with the size of the Galactic bulge.
	Furthermore, from the 511 keV photon flux observed by SPI we conclude that the 511 keV photon production rate in the Galactic center region is $L_{511}=
$7.7$\times$10$^{42}$ ph s$^{-1}$.

	The 511 keV gamma-ray line is produced by $e^{+}$ $e^{-}$ annihilation. 	However before annihilation with an $e^{-}$ of the interstellar medium is possible, the $e^{+}$ must first cool down to the thermal energy of the medium it propagates through.
	Direct $e^{+}$ $e^{-}$ annihilation (with the production of two 511 keV photons per annihilating pair) is possible. 
	However if the temperature is not too high (T$<$10$^6$ K), the formation of an intermediate state, the positronium (Ps), before annihilation is more likely.
	One quarter of the Ps produced are para-prositronium states (with anti-parallel $e^{-}$ and $e^{+}$ spins), their annihilation produces two 511 keV photons, as does the direct $e^{+}$ $e^{-}$ annihilation.
	Three quarters of the Ps (ortho-positronium states, with parallel $e^{+}$ and $e^{-}$ spins) however annihilate with the production of 3 photons whose energy ranges from 0 to 511 keV.
	From the ratio of the fluxes measured for the 511 keV peak and the 3$\gamma$ continuum, the fraction of $e^{+}$ and $e^{-}$ annihilating via the Ps intermediate state can be computed to be $f_{Ps}=8 (6 + 9 \Phi_{511} / \Phi_{3 \gamma})^{-1}$.
	As a result of observations of the Galactic center region with CGRO/OSSE, \cite{kinzer2001} measured a positronium fraction $f_{Ps}=$0.93$\pm$0.04 and a narrow 511 keV line ($<$3 keV FWHM, confirmed by INTEGRAL/SPI).
	This result leaves not much room for direct $e^{+}$ $e^{-}$ annihilation, and shows that the bulk of the positron annihilation from the Galactic center direction occurs after positronium formation in a warm medium, excluding molecular clouds as in \cite{guessum1991} and \cite{ballmoos2003}.

	From the SPI and OSSE measurements the rate of $e^{+}$ annihilation in the Galactic center region can be computed as $L_{e^+}=L_{511}(2 - 3 f_{Ps}/2)^{-1}$. The result is a huge number: every second near the Galactic center $L_{e^+}=$1.3$\times$10$^{43}$ $e^+$ annihilate in a region comparable in size with the Galactic bulge.
	Under the assumption of a steady-state production/annihilation, the same amount of positrons must be produced each second near the Galactic center.
	This conclusion raises the question of the nature of the source capable of injecting such an amount of positrons near the Galactic center, as well as the question of the medium onto which those positrons annihilate, which does not necessarily coincide with the e$^+$ production region, since e$^+$ are transported out of their production site into the medium where they annihilate and which is what we actually observe.

\section{Galactic positron source candidates}

	There are many potential Galactic sources capable of producing emission in the 511 keV gamma-ray line, so the best strategy is to proceed by a process of  elimination.

	Exotic high energy processes in the Galactic center region, such as proton-proton interactions at high energy, neutralino annihilation \citep{bertone2002} and decays of Kaluza-Klein gravitons \citep{hannestad2003, casse2003}, are excluded since the 511 keV emission is not accompanied by a significant flux of high-energy gamma rays which would have been observed by EGRET \citep{hunter1997}.
	However, the annihilation of light dark matter fermions of Majorana type or light dark matter bosons trapped in the Galactic gravitational potential, could produce $e^{+} e^{-}$ pairs at low energy, without high energy counterpart and remain therefore and exciting possibility for the interpretation of the observed 511 keV flux in the Galactic center region, as published in \cite{boehm2004}, presented in these proceedings by \cite{casse2004b} and detailed theoretically by \cite{fayet2004}. 
	In this respect, the search with INTEGRAL/SPI for 511 keV emission in the Sagittarius Dwarf Galaxy, believed to be dark matter dominated, is presented in these proceedings by \cite{cordier2004}.
	In the following, however, we restrict our overview to astrophysical positron source candidates.
	Accreting black holes/microquasars \citep{mirabel1992} could be able to inject positrons in the Galactic bulge, however sources of this kind are not numerous enough and additionally they are active only during short periods of time to be a significant source of Galactic positrons.
	Low Mass X-ray binaries (LMXB) have also been proposed as candidates for positron production.
	The all-sky map obtained by the RXTE survey in the 3$-$20 keV energy band \citep{revnivtsev2004} shows a concentration of sources, among which most are LMXB candidates, located in the central radiant of the Galaxy with a concentration in the Galactic bulge together with a disk component.
	Very recently \cite{lebrun2004} have published a similar distribution of soft gamma-ray sources, detected by INTEGRAL/ISGRI.
	If the positrons produced by those sources annihilate locally, the 511 keV map detected by INTEGRAL/SPI should show a disk component as well, which is not yet the case and disfavors the LMXB hypothesis.

	The most promising remaining sources for Galactic positrons are the radioactive ejecta produced by the various nucleosynthesis sites. 
	Through their $\beta^+$ decays, radioactive isotopes are able to release large amounts of positrons into the interstellar medium, where they annihilate. 
	The $^{22}$Na isotope is expected to be produced by Galactic novae, however their positron production rate is too small and the $^{22}$Na line is absent in the spectrum of the central regions of the Galaxy \citep{leising1998}.
	The $^{26}$Al isotope has been well observed and could produce significant amounts of positrons.
	However the flux of the 1809 keV line of $^{26}$Al in the central Galactic radian, i.e. 3$\times$10$^{-4}$ ph cm$^{-2}$ s$^{-1}$ \citep{knodl1999}, and its narrow latitude profile are incompatible with the detected 511 keV flux and distribution and indicates that $^{26}$Al decay is not the main source of Galactic positrons. The same is true for $^{44}$Ti, with an even smaller expected flux.

	The main source of positrons in the Galaxy is expected to be $^{56}$Co, which decays to $^{56}$Fe, producing a positron with a branching fraction of 19\%.
	$^{56}$Co is the decay product of $^{56}$Ni, which is produced in supernova explosions.
	SNe II produce about 0.1 M$\odot$ of $^{56}$Ni, but have a thick hydrogen envelope which shields the positrons from being released in the surrounding medium.
	On the other hand, SNe~Ia have a thinner envelope and produce a higher amount of $^{56}$Ni, about 0.6 M$\odot$. 
	In their SN~Ia model DD23C, based on the idea that positron escape should modify the late light curve of the supernova, \cite{milne1999} have predicted that 3.3\% of the positrons escape the envelope, reaching the total amount of $\simeq$8$\times$10$^{52}$ $e^+$ released by a typical SN~Ia.
	Under the assumption of a steady-state production/annihilation of positrons, in order to explain the observed positron annihilation rate of $L_{e^+}=$1.3$\times$10$^{43}$ e$^+$ s$^{-1}$ by SN~Ia events alone, a mean SN~Ia rate of 0.6 per century is required in the Galactic bulge.
	As we will see in section \ref{sec_rateSNIa}, this requirement is in excess by a factor of 10 compared to the estimated SN~Ia rate in the Galactic bulge. 
	Therefore we conclude that only a small fraction of SN~Ia events can contribute to the injection of positrons in the Galactic bulge.

	In this article, following \cite{casse2004a}, we would like to advocate an important hypernova contribution to the observed Galactic positron source.
	Hypernovae, i.e. explosions of Wolf-Rayet stars producing strongly asymmetric ejections, have gained attention very recently by the observation of the unusual type Ic supernova SN2003dh linked with the gamma-ray burst GRB030329, as we will discuss in section \ref{sec_hypernova}.

\section{SN~Ia rate in the Galactic bulge}
\label{sec_rateSNIa}

	Stellar photometry observations of stars in the Galactic bulge using 2MASS near infra-red, ESO/NTT and HST data, have permitted to study the Galactic bulge age and metallicity distributions. 
	By summing up the contributions of all individual stars, \cite{zoccali2003} computed the spectral energy distribution (SED) between 500 nm and 2$\mu$m of the Galactic bulge.
	The comparison of the Galactic bulge SED with the SED of other galaxies shows that the Galactic bulge is an elliptical galaxy of type Sa (or Sc) embedded in the Milky Way (which is a spiral galaxy of type Sbc).
	
	Supernova statistics and selection bias studies have led  \cite{cappellaro2003} to evaluate the supernova rate per unit mass in different types of galaxies for the various types of supernovae.
	For a galaxy of type S0a or Sb (like the Galactic bulge), they deduce a SN~Ia rate of $(0.29\pm0.07)\times[10^{11}$ M$\odot$ 10$^{2}$yr$\times($H$/75)^2]^{-1}$.
	The stellar mass of the Galactic bulge is evaluated in a recent synthetic presentation of the structure of the Galaxy by \cite{robin2003} to be $(2.03\pm0.26)\times 10^{10}$M$\odot$, mainly consisting of very old ($\simeq$10 Gyr) stars.
	Taking the recent determination of the Hubble constant H$_0=(71\pm6)$ km s$^{-1}$ Mpc$^{-1}$ by \cite{freedman2001}, we obtain a rate of SN~Ia in the Galactic bulge of 0.05 per century, in good agreement with the value 0.07 obtained by \cite{nomoto2003} and 0.03 by \cite{matteucci1999} through their Galactic evolutionary models.

\section{Hypernova GRB030329/SN2003dh}
\label{sec_hypernova}

	On March 29, 2003, at 11:37:14.67 UT, the HETE-2 satellite detected the gamma-ray burst GRB030329, as reported in \cite{price2003}.
	With a duration of 25 s, this GRB belongs to the category of long duration GRB.
	Its peak flux of $7\times10^{-6}$ erg s$^{-1}$ cm$^{-2}$ in the 30$-$400 keV gamma-ray band rates it among the 0.2\% brightest GRB in the 9 yr BATSE GRB catalog.
	The released isotropic equivalent gamma-ray energy of GRB030329, $0.9\times10^{52}$erg, is comparable to the total gravitational biding energy of a standard core-collapse supernova progenitor star ($\simeq10^{53}$erg), 99\% of which is imparted to neutrinos escaping the supernova, while the kinetic energy of the ejecta is only $\simeq10^{51}$erg, and the energy seen in the optical band is even only $\simeq10^{49}$erg.
	However the gamma-rays of a GRB are not believed to be released isotropically, but produced from internal shocks between expanding shells confined in jets.

	The localization of GRB030329 within an error circle of radius 2 arcmin was available in the GRB Coordinated Network 1.4 h after the event, after which optical telescopes could perform follow-up observations on the optical afterglow  and study the exponential decay of its light curve, as reviewed in \cite{uemura2003}.
	In the first hours, the optical magnitude of the GRB afterglow was $\leq$12, which means that in the first minutes the event could have probably been seen by the naked eye.
	The GRB afterglow emission is believed to be produced by synchrotron emission from the forward shock region in an expanding jet colliding with the ambient medium. 
	A change in the slope of the after-glow could indicated a change of the jet opening angle due to a decrease of its Lorentz boost, and favors models in which the gamma-ray energy is not released isotropically.

	Spectra of the optical afterglow of GRB030329 have been observed at different epochs as reported by \cite{hjorth2003}, and \cite{stanek2003}.
	Early spectra follow a power-law typical for GRB afterglow spectra, superimposed with narrow emission lines from $H_{II}$ regions in the host galaxy.
	A spectral analysis of VTL/UVES data permitted to determine the redshift of the event to be $z=0.1685$, which is equivalent to a luminosity distance of $\sim$800 Mpc (considering the cosmological parameters $\Omega_\Lambda$=0.73, $\Omega_M$=0.27, H$_0$=71~km~s$^{-1}$Mpc$^{-1}$).
	This makes GRB030329 the second closest GRB ever seen, after SN1998bw associated to an atypically weak GRB located at $\sim$37 Mpc.
	As shown by \cite{hjorth2003}, late spectra (acquired after 7 days) are more and more supernova-like and reveal the presence of a supernova event, hereafter named SN2003dh, associated to GRB030329.
	The supernova spectrum is revealed  after subtraction of the underlying afterglow spectrum. 
	It shows the absence of broad H lines, which classifies the supernova as type I.
	Absent He and Si lines and the appearance of broad features identified with Mg and O (after 27 days) indicate a SN~Ic, an explosion of a Wolf-Rayet star, which is a C+O core which has lost its envelope.
	From the broadening of the spectral lines, \cite{hjorth2003} conclude that the expansion velocities are very high, up to 3.6$\times$10$^4$ km s$^{-1}$ about 10 days after the explosion.
	Furthermore they show that the time-scale of the supernova light curve is unusually short (compared to a typical SN~Ic, as e.g. SN1998bw) and conclude that the event is a hypernova, an axial-symmetric supernova explosion of a Wolf-Rayet star seen on axis.

\begin{figure}
\centering
\includegraphics[width=0.9\linewidth]{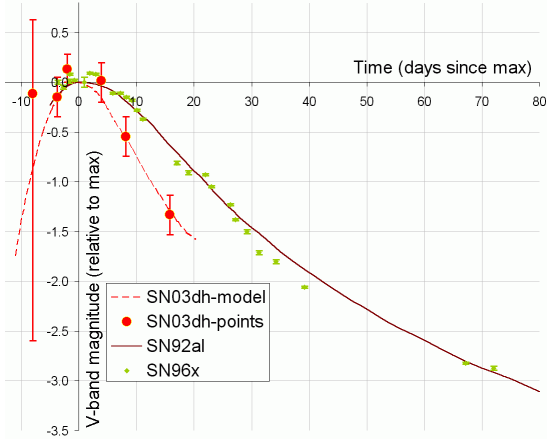}
\caption{
Comparison of the optical V-band light curve of SN2003dh \citep{hjorth2003} with V-band light curves of two standard SN~Ia, namely SN1996X \citep{salvo2000} and SN1992al \citep{hamuy1996}.
For each SN the V-band light curve has been normalized to its maximum (t=0, magnitude=0).
\label{fig_lightcurve}}
\end{figure}

	In Figure~\ref{fig_lightcurve} we compare the light curve of SN2003dh obtained by \cite{hjorth2003} with standard SN~Ia light curves.
	The decrease of supernova optical light curves is known to follow the radioactive decay of $^{56}$Co, the daughter nucleus of $^{56}$Ni, synthesized by the supernova explosion. 
	Therefore the supernova ejecta must be powered by the $^{56}$Co radioactivity, namely its decay products, the 847 keV $\gamma$-photon and the $e^{+}$ (emitted 19\% of the times, with an energy peaking at 640 keV and a maximum energy at 1.4 MeV) produced by $\beta^+$ decay.
	At early times the ejecta are opaque to the $\gamma$ and the $e^{+}$, which enables them to heat the ejecta and to power the optical light curve.

	In SN2003dh we observe a much faster decrease of the light curve than in a typical SN~Ia.
	From this we conclude that the ejecta must be thinning out faster. 
	Therefore they are heated-up less efficiently by the $^{56}$Co radioactivity, consequently its decay products, the $\gamma$ and the $e^{+}$,  must be leaking out before they can heat up the ejecta.
	The yield of released positrons should therefore be increased.
	From this simple light-curve comparison we conclude that SN2003dh must release more positrons than a SN~Ia for a comparable amount of $^{56}$Ni initially synthesized in the explosion.

\section{SN~Ia positron yield}\label{sect_snIa}

	Positron escape from SN~Ia has been studied by \cite{milne1999} and \cite{milne2001}, who compared different model predictions with optical V-band light curves of 22 supernovae of type Ia.
	A good fit is obtained with the delayed-detonation model DD23C, who gives the following parameters for a typical SN~Ia.
	The mass of the ejecta is $M$=1.34~M$\odot$, the total mass of the white dwarf (Chandrasekar mass) in the progenitor binary system being ejected in the explosion.
	The kinetic energy of the ejecta is $E=\frac{1}{2} M v^2$=1.17$\times$10$^{51}$~erg. 
	The explosion synthesizes a mass $M_{Ni}$=0.6 M$\odot$ of $^{56}$Ni (among which 3.3\% remain when the envelope becomes transparent to positrons) and eventually releases $N_{e^+}$=8$\times$10$^{52}$ positrons into the interstellar medium.

	Taking those numbers as an input, we compute the leak-out time of the positrons using a simple spherical symmetric explosion model, in which we assume a free homologous expansion of the sphere containing the ejecta (whose radius   follows at time $t$ the relation $R = v t$, where $v$ is the ejecta velocity, determined by the kinetic energy of the explosion $E=\frac{1}{2} M v^2$).
	We consider that the synthesized radioactive nuclei ($^{56}Ni$) remain confined deeply inside the ejecta and that positrons are produced in the decay chain $^{56}$Ni$\to ^{56}$Co (with a half-life of 6.1 days) followed by $^{56}$Co$\to ^{56}$Fe (with a half-life of 77 days and releasing a e$^+$ with a branching fraction $Br$=19\%).
	We assume a purely absorptive positron propagation through the ejecta. 
	Positrons escape the ejecta when the column density traversed be positrons $\rho R$ drops below their mean free path $\lambda$ inside this medium, therefore the positron leak out time scales as $t_{e^+} \propto M E^{-1/2}$.

	Within this model, in order to release $N_{e+}$=8$\times$10$^{52}$ $e^+$ out of $M_{Ni}$=0.6 M$\odot$ of $^{56}$Ni, we compute transparency to positrons to be reached $t_{e^+}$=390 d after the explosion, which is consistent with the SN~Ia late light curve fit using the model DD23C including positron radial escape, Figure 3 in \cite{milne2001}.
	The transparency to positrons is reached when the column density $\rho R$ drops below $\lambda$=0.6 g cm$^{-2}$, corresponding to the mean free path of MeV electrons propagating through C or O as measured at CEA-Saclay by \cite{pages1970}.
	At this moment, only 3.3\% of the $^{56}$Co mass has not yet decayed, in agreement with model DD23C.
	Only the positrons formed in the decay of those remaining $^{56}$Co nuclei are not trapped inside the ejecta but escape into the interstellar medium.

\section{Hypernova positron yield}

	The light curve of SN2003dh has been reproduced by a two component model by \cite{woosley2003} where the asymmetry of the star explosion is modelled by a fast polar combined with a slow equatorial ejection, similar to \cite{maeda2003}.
	In this model, a 10 M$\odot$ rotating Wolf-Rayet star (a bare C+O core which lost its envelope) explodes and synthesizes $M_{Ni}$=0.5 M$\odot$ of $^{56}$Ni; it produces 8 M$\odot$ of ejecta and a compact remnant of 2 M$\odot$.
	Within a cone, whose half opening angle is 45$^\circ$, $M$=1.2 M$\odot$ are ejected (per cone) with a high kinetic energy $E=$1.25$\times$10$^{52}$ erg;  the remaining 5.6 M$\odot$ are ejected into the equatorial region with slow velocity.

	Within this model, the calculation for the symmetric case (section \ref{sect_snIa}) holds per cone.
	The positron leak-out time scales like $t_{e^+} \propto M E^{-1/2}$; the column density $\rho R$ crosses the positron mean free path limit $\lambda$=0.6 g cm$^{-2}$ at which the ejecta become transparent to positrons at the time $t_{e^+}$=107 d; at this time 42\% of the $^{56}$Co mass has not decayed and will produce 10$^{54}$ positrons which escape the ejecta in the case where the radioactive $^{56}$Co remains confined deeply inside the ejecta.
	Moreover, in the case of strong mixing of $^{56}$Co with the ejecta, indicated by the short rise-time of the light curve, virtually all the 2$\times$10$^{54}$ e$^+$ formed by the $^{56}$Co decay escape.
	We consider therefore that hypernovae of the kind of SN2003dh could contribute significantly to the cosmic positron injectors.
	In particular SN2003dh would produce 25 times more e$^+$ than a supernova of type Ia.
	The positron annihilation rate observed by SPI would be explained by SN2003dh-like hypernovae alone, if their average occurrence is of the order of 0.02 per century.

	The hypernova rate in the Galactic center region is very uncertain, however a first crude estimate can be computed in the same way as we have done for the SN~Ia rate in the Galactic bulge.
	Considering the Galactic nuclear zone as a spiral galaxy (embedded inside 
the elliptical Galactic bulge) with a stellar mass content of 6.3$\times$10$^{9}$ M$\odot$, as given in \cite{gusten2004}, and knowing the SN Ib\&Ic rate in a spiral galaxy to be 0.30$\times$[10$^{11}$ M$\odot$ 10$^{-2}$ yr $\times$ (H/75)$^2$]$^{-1}$ from \cite{cappellaro2003}, the global SN Ib\&Ic rate in the Galactic nuclear zone becomes 0.02 per century.
	From this value, we try now to get a first very crude hypernova rate estimate for the Galactic nuclear zone.
	According to the very recent estimate of \cite{podsiadlowsk2004}, most SN Ib\&Ic actually appear to belong to the SN Ic sub-type and only a fraction of 5\% of {\it observed} SNe Ic are hypernovae, but observational biases due to the over-luminosity of hypernovae could decrease this fraction to about 1\%.
	Furthermore, out of the 5 so far observed hypernovae (SN1997dq, SN1997ef, SN1998bw, SN2002ap, SN2003dh), one (SN2003dh) is a favorable candidate for positron production.
	However, the fraction of binary systems in the dense stellar clusters located in the Galactic nuclear zone is expected to be exceptionally high \citep{portegies2004}, favoring asymmetric hypernova explosions, which could be a factor 2 more frequent than on average in the Galaxy.
	Taking these numbers at face value, a first crude estimate of the rate of  SN2003dh-like hypernovae in the Galactic nuclear zone becomes about 10$^{-4}$ per century, which would be unfavorable for hypernovae being a dominant positron injection candidate.
	Much work however remains to be done to refine this first very crude estimate.


	The supernova remnant (SNR) Sgr A East, recently studied by \cite{maeda2002} using X-ray data from Chandra, is a very promising hypernova candidate in the Galactic nuclear zone.
	This SNR is located in the inner few pc of the Galaxy, and is the remnant of the core collapse of a massive star (with a mass between 13 and 20 M$\odot$) about 10$^4$ years ago.
	Furthermore, this remnant has been related to the EGRET source 3EG J1746-2852 by \cite{fatuzzo2003}.
	The EGRET high $\gamma$-ray luminosity could be produced by the interaction of the expanding SNR in the surrounding high density (up to 10$^3$ cm$^{-3}$) and highly magnetized (up to 0.2 mG) medium, however the accompanying positron source would look rather point-like to SPI, which is excluded.

\begin{figure}
\centering
\includegraphics[width=0.9\linewidth]{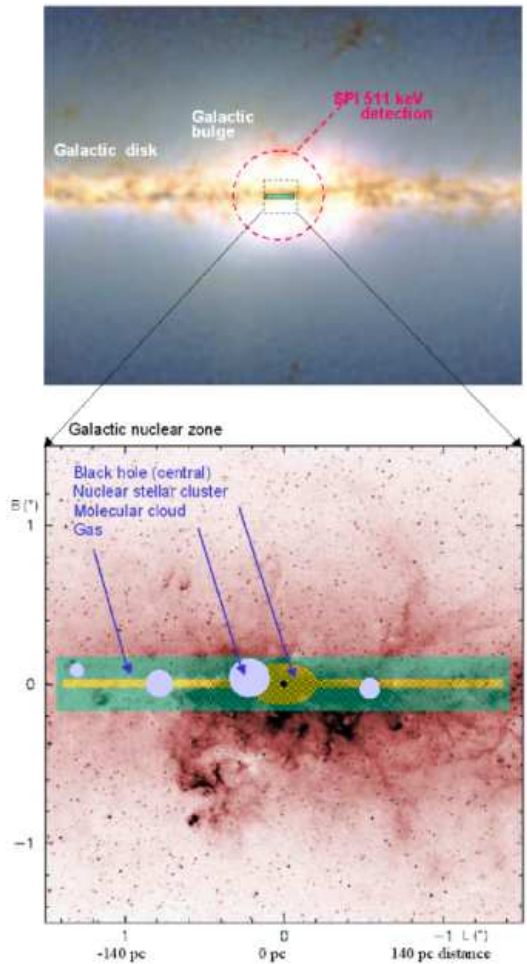}
\caption{
	The upper figure shows the Galactic center region compiled from star counts in the Two Micron All Sky Survey (2MASS) database of about 100 million stars, the colors representing the local density of stars seen in survey (image credit: 2MASS/J. Carpenter, M. Skrutskie, R. Hurt, http://www.ipac.caltech.edu/2mass).
	The Galactic disk and bulge (with a lateral extension of about 12~$^\circ$) appear clearly, and the 511 keV gamma-ray emission region observed by INTEGRAL/SPI has been sketched (with a FWHM of about 10~$^\circ$).
	In the lower figure a zoom of the Galactic nuclear zone is shown, overlaid with the infrared 8 $\mu$m image of the region observed by MSX \citep{bland2003}.
	Also sketched are the locations of the central black hole, the nuclear stellar cluster and disk, as well as the giant molecular clouds, observed in \cite{kim2002}, surrounded by the gas in the Galactic nuclear zone.
\label{fig_galcenter}}
\end{figure}

\section{Positron propagation in the Galactic center region}

	In this section, we focus on the Galactic center region (summarized in Fig. \ref{fig_galcenter}), where hypernovae are likely to occur and release the positrons they produce.
	The large-scale physical characteristics of stars and interstellar matter in the nuclear zone of the Galaxy have been described recently by \cite{launhardt2002}.
	While the sun is located in the thin Galactic disk, containing also compact Giant Molecular Clouds, the Galactic bulge with a few kpc in diameter is composed of old stars embedded in a low density gas, in which we do not expect many core collapse supernovae to occur, and hence no hypernovae.
	On the other hand, the nuclear zone of the Galaxy is a very active zone, containing many young stars, suitable as core collapse supernova and hypernova progenitors.
	The nuclear zone has a bar-like geometrical shape with 450 pc in diameter and a height of 50 pc, as described by \cite{gusten2004}.
	The stellar mass content of the nuclear zone amounts to $6.3\times10^{9}$ M$\odot$, emitting a luminosity of $2.5\times10^{9}$ L$\odot$. 
	About 70\% of this luminosity is produced by young massive main sequence stars in the optical and UV bands. 
	Those stars form a nuclear stellar cluster at the center (with a density decreasing as $R^{-2}$) and are surrounded by a nuclear stellar disk which extends up to R$\simeq$225 pc giving this star collection a spiral galaxy like shape. 
	Inside the Galactic nuclear zone are also located huge molecular clouds (with a mass of $8\times10^{7}$ M$\odot$), which are very massive concentrations of non-stellar matter and are probably prevented from collapse by tidal force effects of the central black hole, which amounts to a mass of about 2$\times 10^6$ M$\odot$.
	The molecular clouds distribution is very clumpy; the clouds have a very low filling factor of only about 1\% and are made of a high density gas (of about 10$^4$ cm$^{-3}$).
	The CO J=4$\to$3 transitions, for which maps have recently been published by \cite{kim2002} in the Galactic nuclear zone, based on data from the AST/RO sub-millimeter range telescope located close to the south pole, are good tracers of molecular hydrogen concentrations and permit to detect the shape and location of the molecular clouds in the Galactic nuclear zone.
	The mass of gas around the molecular clouds amounts to $2.5\times10^{6}$ M$\odot$ and is distributed in an inner cold disk with a temperature of $\sim$30 K which extends up to R$\simeq$110 pc, and an outer even colder torus which makes up about 80\% of the mass with a temperature of $\sim$20 K and a radial extension up to R$\simeq$225~pc.

	At this point we would like to address the question why we do not expect to see positron annihilation from the Galactic disk, but only the from the Galactic center, in the hypothesis of positrons produced by hypernovae.
	Indeed, hypernovae are also expected to explode in the Galactic disk, however positrons produced by their explosion in the Galactic center region are more likely to be confined in this region, which is embedded in the Galactic bulge, offering confinement and matter for positrons to annihilate.
	On the contrary, positrons produced by hypernova explosions in the Galactic disk are likely to leak out of the very thin zone of the Galactic disk ($\sim$100 pc in height), and could escape the Galaxy without in-situ annihilation.
	Furthermore, in the dense star forming region of the Galactic nuclear zone the formation of binary systems is more frequent, which favors asymmetric hypernova explosions.
	Additionally, the initial mass function of the stars of mass $m$ in the Galactic disk is $dN / dm \propto m^{-3}$ according to \cite{robin2003} (with a slope $\alpha$=3 a bit above than the Salpeter assumption $\alpha$=2.35), while it is much less steep in the Galactic nuclear zone, where the relation is $dN / dm \propto m^{-\alpha}$ with $\alpha$ between 1.2 and 1.6 according to \cite{figer2004}.
	Consequently, even if the Galactic disk contains more massive stars than the Galactic nuclear zone, the relation could be more balanced for very massive stars, more likely to be hypernova progenitors.
	Finally, taking into account the geometry, the production of positrons per solid angle is much higher in the Galactic center region than in the Galactic disk, in agreement with the observations.

	If we suppose that positrons are produced by hypernova explosions of young stars populating the Galactic nuclear zone, let us focus in the rest of this section on the propagation and possible annihilation of those positrons in the nuclear zone. 
	
	We consider first the case of positron interactions with the molecular clouds in the zone.
	According to \cite{gusten2004}, the temperature of the gas in the clouds is between $\sim$50 and 70 K, which is a temperature below the charge exchange threshold. 
	Therefore, no positronium formation is expected for positrons penetrating inside the clouds.
	Since we observe that the bulk of positron annihilation occurs after positronium formation, the annihilation site of the positrons can not be the molecular clouds.
	Furthermore, the density inside the clouds is $n\simeq 10^4$ cm$^{-3}$.
	Let us apply the energy loss formula per unit time expressed in the units eV s$^{-1}$, as published by \cite{longair1981}, see also \cite{boehm2004}, i.e. $dE/dt = -2\times 10^{-8} n [6.6+\ln(E/M)]$, where $E$ and $M$=511 keV are the energy and mass of a positrons trying to penetrate the clouds of density $n$ expressed in cm$^{-3}$. 	
	We compute that the stopping time of a positron of typically $E=$1.2 MeV produced by $^{56}$Co decay (with a kinetic energy peaking at 640 keV) is $t \simeq$7 yr and therefore the penetration depth of a positron is limited to a few pc only.
 	Consequently supernova positrons penetrate only a thin shell surrounding the huge molecular clouds.

	The bulk of the mass located inside the clouds is therefore unreachable for the positrons, which mainly propagate in the gas surrounding the clouds, since the mass contained in the shells of the clouds is small compared to the mass of the gas around the clouds.
	The density of the gas between the molecular clouds can be evaluated to be $n\simeq 14$ cm$^{-3}$ from the values of the mass of the gas ($2.5\times10^{6}$ M$\odot$) and the volume of the nuclear zone, as given by \cite{gusten2004}.
	In this gas, the stopping time of positrons can be evaluated to be $t\simeq$5200 yr.
	From this value, we conclude that a steady-state positron production is possible if hypernovae explode in the nuclear zone as we have previously evaluated every $\sim$5000 years on average, and if their released positrons propagate in the gas around the molecular clouds.

	Positrons produced in the nuclear zone by hypernova explosions could leak out of their production zone, into the very low-density gas of the Galactic bulge.
	In this medium the positron stopping time would become much longer and they could accumulate and fill-up eventually the whole Galactic bulge, where they would annihilate.
	Mechanisms of propagation of matter out of the nuclear zone exist: non thermal filaments have been detected, see \cite{laroza2004}, the existence of a bipolar magnetic field orthogonal to the galactic plane has been established, see \cite{rodriguez2004}.
	An infra-red map of the Galactic bulge, as seen by MSX in the 8~$\mu$m band  and published by \cite{bland2003}, reveals the presence of dust covering a zone a bit larger in size than the Galactic nuclear zone.
	The presence of dust could be accompanied by the presence of a warm ($\sim$10$^4$ K) and low density gas, with which positrons could form a significant fraction of positronium before annihilation, and produce a narrow 511 keV line. 
	Therefore we propose to study whether the MSX 8$\mu$m infra-red map could be compatible with the map of the 511 keV annihilation region seen by INTEGRAL/SPI, in which case a candidate positron annihilation site would have been found.
	However, in this context, we would like to mention that the extension of the dust tracing infra-red map is about 2$^\circ$, while the extension of the 511 keV annihilation region is about 9$^\circ$ in size as seen by SPI (according to \cite{jean2003} between 6$^\circ$ and 19$^\circ$ FWHM at 95\% C.L.), marginally compatible with OSEE (\cite{purcell1997}, refined by \cite{kinzer2001}: 4.9$^\circ$$\pm$0.7$^\circ$ FWHM in latitude and 6.3$^\circ$$\pm$1.5$^\circ$ FWHM in longitude).

	Furthermore, positrons could be disseminated at much greater distances into the Galactic bulge, by the effect of a bipolar wind detected at large scales, as discussed in \cite{bland2003}.
	In this case, positrons could fill the entire Galactic bulge and the positron annihilation map produced by SPI would not be able to pinpoint the location of the positron source, but would rather give the distribution of matter on which the annihilation takes place; positrons produced by hypernovae in the Galactic center region or by decaying dark matter particles in a wider region would propagate in the whole Galactic bulge and both possibilities could not be distinguishable by the annihilation map.

	Finally, we would like to mention the possibility that positrons could be produced in a non-steady manner. 
	Indeed, as shown in \cite{bland2003}, a starburst activity took place about 7$\times$10$^6$ years ago in the Galactic center region (and may be recurrent on longer time scales).
	Positrons could have been formed in this starburst, through SN2003dh-like hypernova events, and disseminated in the Galactic bulge, where they are still present and annihilate at a rate related to the properties of the annihilation medium they encounter.
	We adopt the estimate of \cite{bland2003}, that the number of supernovae in this burst is 10$^4/\epsilon$, where $\epsilon$ is the thermalization efficiency of the supernova ejecta.
	Depending on the density of the medium, $\epsilon$ can be less than 10\% \citep{thorton1998}. 
	Taking $\epsilon$=0.1 for illustrative purpose, one estimates that 10$^5$ SN exploded 7$\times$10$^4$ centuries ago, leading to 1.6$\times$10$^4$ SN~Ic, among which 160 are hypernovae, and 60 are SN2003dh-like (taking binarity into account).
	This corresponds to a rate of about 10$^{-3}$ century$^{-1}$, which alleviates the problem, but is still low, although this is admittedly just an order of magnitude estimate.
	Note that this line of reasoning is only valid if the stopping time of positrons is larger than 7 Myr, which requires a density of the medium lower than 10$^{-2}$ cm$^{-2}$.
	Furthermore, the ancient starburst hypothesis could explain, why $^{26}$Al, which would also have been produced in the starburst, is not seen anymore; it would simply have decayed since its formation, due to its relatively smaller half life-time of about 0.7 million years.

\section{Concluding remarks}

	INTEGRAL/SPI has recently detected positron annihilation in the Galactic center region, which is compatible in size with the Galactic bulge, and whose rate is of the order of 1.3$\times$10$^{43}$ e$^+$ s$^{-1}$.
	Prompted by the claim that those positrons could be produced by the annihilation of a new kind of light dark matter particles in the Galactic center region, we have studied astrophysical candidate sources which could be capable of injecting the observed amounts of positrons in the Galactic center region.
	We have ruled out SNe~Ia as the dominant positron injectors, due to their too low explosion rate in the Galactic bulge.
	We have proposed an alternative solution, namely hypernovae, capable of injecting up to 25 times more positrons than a typical SN~Ia. 
	Hypernovae are likely to occur in the Galactic center region, from which the produced positrons could escape and fill up the entire Galactic bulge.
	The rate of hypernova events remains uncertain, but first crude estimates show that it might also be too low to be compatible with the steady-state positron production/annihilation hypothesis.
	The observed positrons could therefore be the remains of a starburst in the Galactic center region, which occurred a few million years ago.
	Before excluding hypernovae as possible positron injectors in the Galactic center region, more observational constraints are required.


\section*{Acknowledgments}

	The authors would like to thank their colleagues of SPI, the Spectrometer aboard the European Space Agency's gamma-ray space telescope INTEGRAL, for fruitful discussions.


\end{document}